# La-substituted CaFeAsH superconductor with $T_c$ = 47 K

Yoshinori Muraba[1], Satoru Matsuishi[2], Hideo Hosono[1,2,3,*]

[1]*Materials and Structures Laboratory, Tokyo Institute of Technology, 4259 Nagatsuta-cho, Midori-ku, Yokohama 226-8503, Japan*
[2]*Materials Research Center for Element Strategy, Tokyo Institute of Technology, 4259 Nagatsuta-cho, Midori-ku, Yokohama 226-8503, Japan*
[3]*Frontier Research Center, Tokyo Institute of Technology, 4259 Nagatsuta-cho, Midori-ku, Yokohama 226-8503, Japan*

La-substituted CaFeAsH, $(Ca_{1-x}La_x)FeAsH$, were synthesized by the solid state reaction at 1173 K under high pressure of 2.5 GPa. The $\rho-T$ anomaly corresponding to tetragonal to orthorhombic structural transition was suppressed by La-substitution and then the superconductivity was observed at an electron doping content of $N_e > 0.05$. The maximum onset $T_c$ of 47.4 K was reached for $N_e = 0.18$, which is significant increase compared to 23 K achieved by direct electron doping to the FeAs-layer via Co-substitution to the Fe-site. This value of $T_c$ value fits well to the phenomenological relation between $T_c$ and the bond angle of As-Fe-As in indirectly electron doped $Ln$FeAsO ($Ln$ = Lanthanide). These results indicate that the electron was doped via the indirect mode through the aliovalent ion substitution to CaH-layer ($Ca^{2+} \rightarrow La^{3+} + e^-$).

Since the discovery of superconductivity in LaFeAsO$_{1-x}$F$_x$ with $T_c$ = 26 K,[1] various types of FeAs-based and related layered compounds have been reported as candidates for high $T_c$ superconductors.[2–11] $T_c$ is raised up to 55 K in SmFeAsO$_{1-x}$F$_x$ as a consequence of these efforts.[12–16] The 1111-type $Ln$FeAsO ($Ln$ = lanthanide) is a parent compound for iron arsenide superconductors with ZrCuSiAs-type structure (space group $P4/nmm$) composed of FeAs anti-fluorite-type conducting layers sandwiched by $Ln$O fluorine-type insulating layers. They undergo a structural transition from the tetragonal to the orthorhombic phase, accompanied by a transition from a paramagnetic to a stripe-type antiferromagnetic phase between at 140-180 K.[17,18] These transitions are suppressed by carrier doping via partial element substitution and then the superconductivity is induced. The carrier doping mode for 1111-type iron arsenides is classified into two types, "indirect doping" and "direct doping" depending on the substituted layer. The former is aliovalent substitution in the $Ln$O layer such as fluorine-substitution to the oxygen site (O$^{2-}$ = F$^-$ + e$^-$) and the latter is element substitutions in the FeAs layer such as cobalt or nickel substitution to the iron site. The indirect doping leads to a higher $T_c$ than direct doping because the structural and electrical disturbances in the superconducting FeAs layer are less, e.g. 55K for SmFeAsO$_{1-x}$F$_x$[16] vs. 17K for SmFe$_{1-x}$Co$_x$AsO.[19] In addition to $Ln$FeAsO, $Ae$FeAsF ($Ae$ = Ca, Sr, Eu) and CaFeAsH are new parent compounds that belong to the 1111-type iron arsenides containing $Ae$F or CaH layers as a building blocks.[20–23] By the direct electron doping, these compounds exhibit the highest $T_c$ ~ 23 K among Co-substituted 1111-type iron arsenides.[20,23] Therefore, the indirect doping to these compounds is expected to be a promising pathway to renew the $T_c$ record of iron-based superconductors. Although the indirect doping to $Ae$FeAsF by $Ln$-substitution to the Ca-site was already reported in earlier papers,[24–26] a notable amount of impurity phase containing $Ln$ segregated in the samples complicates any conclusive evidence for $Ln^{3+}$-substitution to the Ca$^{2+}$ site, e.g. peak shift and change in peak intensity ratio of XRD due to $Ln$-substitution were not confirmed. In our previous works, we reported that a large amount of H$^-$ ion can substitute for oxygen in $Ln$FeAsO instead of F$^-$ by using high pressure synthesis.[27,28] In this study, we performed the indirect electron doping to 1111-type CaFeAsH by La-substitution of the Ca-site and confirmed the bulk superconductivity with maximum $T_c$ = 47 K.

Ca$_{1-x}$La$_x$FeAsH was synthesized by the solid-state reaction of LaH$_2$, CaH$_2$ and FeAs,

$$(1-x)\ CaH_2 + x\ LaH_2 + FeAs \rightarrow Ca_{1-x}La_xFeAsH + 1/2H_2\uparrow,$$

using a belt-type high pressure anvil cell. CaH$_2$, LaH$_2$ and FeAs were prepared from their respective metals, and CaH$_2$ or LaH$_2$ was synthesized by heating Ca or La metal in a H$_2$ atmosphere. All starting materials and precursors for the synthesis were treated in a glove box filled with purified Ar gas ($H_2O$, $O_2$ < 1 ppm). The mixture of starting materials was placed in a BN capsule with a mixture of $Ca(OH)_2$ and $NaBH_4$ as an excess hydrogen source. This hydrogen source generates the $H_2$ via following reaction accompanying decomposition of $Ca(OH)_2$ to CaO and $H_2O$ at ~573 K,

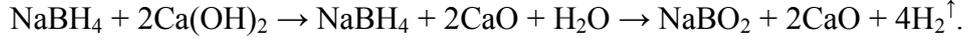
$$NaBH_4 + 2Ca(OH)_2 \rightarrow NaBH_4 + 2CaO + H_2O \rightarrow NaBO_2 + 2CaO + 4H_2\uparrow.$$

A BN capsule was put in a stainless capsule with a stainless cap to isolate the starting materials from an external atmosphere. The capsules were heated at 1173 K and 2.5 GPa for 30 min. The crystal structure of the resulting materials was examined by powder X-ray diffraction (XRD: Bruker D8 Advance TXS) using Cu $K_\alpha$ radiation with the aid of Rietveld refinement using Code TOPAS4.[29] The elemental composition except hydrogen was determined by an electron-probe micro analyzer (EPMA; JEOL Inc. model JXA-8530F) equipped with a field-emission-type electron gun and wavelength dispersive x-ray detectors. The micrometer-scale compositions within the main phase were probed on ten focal points and results were averaged. The amount of hydrogen incorporated in the samples was measured by thermal desorption spectrometry (TDS; ESCO Inc., TDS-1000 S/W). The dc resistivity ($\rho$) and magnetic susceptibility ($\chi$) were measured in the temperature range of 2 – 200 K, using a physical properties measurement system (Quantum Design Inc.) with a vibrating sample magnetometer attachment. $\chi$ was measured on bulk and powdered samples and the data of the powder were adopted.

Figure 1 (a) shows XRD patterns of $Ca_{1-x}La_xFeAsH$ with nominal $x$ ($x_{nom}$) = 0.1−0.4. For $x_{nom}$ up to 0.3, major peaks were indexed as reflections from a ZrCuSiAs-type phase with $P4/nmm$ space group. Some minor peaks were identified as reflections from BN (1.9 wt. %) and $CaFe_2As_2$ (5.1 wt. %) for $x_{nom}$ = 0.1 and LaAs (0.5 wt. %) for $x_{nom}$ = 0.3. The LaAs impurity content increased up to 8.5 wt. % at $x_{nom}$ = 0.4, indicating that the solubility limit of La in CaFeAsH was $x$ ~ 0.4. With increase of $x_{nom}$, the 012 reflection of $Ca_{1-x}La_xFeAsH$ at $2\theta$ ~ 30° was enhanced and exceeded the 001 reflection at $2\theta$ ~ 10°. This observation is understood as an effect of increase in atomic scattering factor of the Ca-site, *i.e.*, La-substitution to the Ca-site. Figure 1(b) shows the lattice parameters as a function of $x_{nom}$. Such a lattice expansion is natural since the ionic radius of $La^{3+}$ (116 ppm) is larger than that of $Ca^{2+}$ (112 ppm). Rietveld refinement was performed with a structure model in which La replaces Ca sites in the 1111-type structure. The values of the temperature factors ($B_{eq}$) for Ca, Fe and As were fixed at those in CaFeAsF,[30] and the values of $B_{eq}$ for La and O were fixed at those of LaFeAsO.[18] Figure 1(c) shows the site occupancy of $La^{3+}$ in the $Ca^{2+}$ site as a function of $x_{nom}$. Those results indicate that $La^{3+}$ occupies the $Ca^{2+}$ site and the site occupancy is comparable to $x_{nom}$. Figure 1 (d) shows the As-Fe-As bond angle α as a function of $x_{nom}$, which slightly increased with $x_{nom}$ to approach the regular tetrahedron angle (~109.5°).

Figure 2 (a) shows La concentration $x$ analyzed by EPMA and normalized by the molar content of iron in $Ca_{1-x}La_xFeAsH$ as function of $x_{nom}$. The analyzed $x$ is proportional to $x_{nom}$ and its standard deviation increases with $x_{nom}$. Figure 2 (b) shows oxygen concentration analyzed by EPMA and normalized by the molar content of iron (smaller blue closed circle and line) and hydrogen concentration determined by TDS (green closed triangle and line). The oxygen concentration is 0.06 at $x = 0$ and increases up to 0.10 with $x$. The hydrogen concentration is 1.00 at $x = 0$ and decreases down to 0.94 with increasing $x$. This deviation occurs only in the 1111-type main phase because there are no impurity phases bearing hydrogen in $Ca_{1-x}La_xFeAsH$. The analysis of powder neutron diffraction pattern of CaFeAsH verified that hydrogen fully occupies the anion site in the block layer.[31] Therefore, the oxygen observed by EPMA in $x = 0$ mainly derived from surface contamination during sample preparation procedures for EPMA analysis (probed depth of EPMA operation at 10kV is ~3 μm) such as surface polishing. By assuming the amount of surface oxygen is constant among all samples, the oxygen content in the bulk was estimated by subtracting the oxygen content at $x_{nom} = 0$ from the observed oxygen content ($y$: blue closed square and line in Fig 2(b)). The total amount of hydrogen and internal oxygen, $x + y$, is always close to unity, implying that the hydrogen site in the CaH blocking layer is partially occupied by oxygen. The oxygen in the hydrogen site should act as an electron acceptor ($H^- = O^{2-} + h^+$), in contrast to $La^{3+}$ substituting the $Ca^{2+}$ site that acts as electron donor ($Ca^{2+} = La^{3+} + e^-$). Therefore, the total number of doped electron per iron ($N_e$) can be expressed as $x - y$.

Figure 3(a) shows the temperature dependence of the electrical resistivity ($\rho$) in $Ca_{1-x}La_xFeAsH$ with $x = 0, 0.08, 0.23$ and $0.33$. For $x = 0$, an anomaly due to the structural and the magnetic transitions was seen around ~100 K. As $x$ is increased, the anomaly was suppressed (small kink remain around 100 K in $x = 0.08$) and zero resistivity was observed. The resistivity determined onset $T_c$ ($T_c^{onset}$) remained almost constant (47.4 K) and $\Delta T_c$ were ~10 K ($x = 0.08$), 6 K ($x = 0.23$), 10 K ($x = 0.33$), respectively. The exponent $n$, which is obtained by fitting the data to $\rho(T) = \rho_0 + AT^n$ ($\rho_0$: residual component of resistivity) in the temperature region between $T$ just above $T_c$ and 95 K, decrease with increasing $x$. The temperature dependent electrical resistivity of the normal conducting state changed even though $T_c^{onset}$ remains nearly unchanged with $x$. Non-Fermi liquid behavior ($n \sim 1$) is observed in the entire doped region. Figure 3(a) shows magnetic susceptibility ($4\pi\chi$) vs $T$ plots of samples with $x = 0.08, 0.23$ and $0.33$ under zero field cooled (ZFC) and field cooled (FC) with a magnetic field of 10 Oe. The diamagnetism due to superconductivity was clearly observed below $T_c^{mag}$, 28.2 K ($x$ = 0.08), 40.6 K ($x$ = 0.23) and 35.7 K ($x$ = 0.33) and the values of $-4\pi\chi$ over 0.66 at 10 K show bulk superconductivity.

Figure 4(a) summarizes the electronic phase diagram of $Ca_{1-x}La_xFeAsH_{1-y}O_y$ as function of the number of doped electron ($N_e = x - y$), superimposed on the data of $CaFe_{1-x}Co_xAsH$.[23] The maximum $T_c^{onset}$ of 47.4 K at $N_e$ = 0.16 is almost double as high as that of Co-substitution ($T_c$ = 23 K). Although $T_c^{onset}$ remains constant with $N_e$, $T_c^{mag}$ shows a dome-like structure. The difference between $T_c^{onset}$ and $T_c^{mag}$ indicates the inhomogeneity of $N_e$ within the samples i.e., $T_c^{onset}$ is dominated by domains with a maximum $T_c$.

Figure 4(b) shows $T_c$ of indirectly electron doped 1111-type iron arsenides as a function of $\alpha$,[27,31–38] the so called Lee-plot.[39] The $T_c$ increases with change in $Ln$ from La to Sm, reaches the maximum in $Ln$ = Sm or Gd (where α is near the regular tetrahedron angle of ~109.5°), and then decreases for $Ln$ = Tb, Dy. The maximum superconducting transition temperature, $T_c^{max}$ = 47 K in $Ca_{1-x}La_xFeAsH$ is comparable to that of indirectly electron doped $Ln$FeAsO. The present material falls well on the master curve of the Lee-plot, indicating that superconductivity in $Ca_{1-x}La_xFeAsH$ is caused by indirect electron doping.

Here we discuss the possibility that the superconductivity observed here was caused by $Ca_{1-x}La_xFeAs_2$ ($T_c$ = 45 K)[40,41] or $Ca_{1-x}La_xFe_2As_2$ ($T_c$ = 47 K)[42] which were contained as impurity phase in the $Ca_{1-x}La_xFeAsH$ because their constituent elements and $T_c$ are similar to those of $Ca_{1-x}La_xFeAsH$. First, we consider the possibility of superconductivity derived from $Ca_{1-x}La_xFeAs_2$. The results of Rietveld analysis indicate that the $Ca_{1-x}La_xFeAs_2$ content was 0.04 wt. % or less. It is thus impossible that such a tiny amount of $Ca_{1-x}La_xFeAs_2$ yields the large values of $-4\pi\chi$ over 0.66. Next considered is the possibility of superconductivity derived from $Ca_{1-x}La_xFe_2As_2$. It is reported that $Ca_{1-x}La_xFe_2As_2$ contain two superconducting phases; the higher $T_c$ phase ($T_c$ = 47 K) shows filamentary superconductivity and the lower $T_c$ ($T_c$ = 20 K) phase does bulk superconductivity. The observed ZFC and FC curves clearly separated at over 20 K and the superconductivity observed at > 20 K (values of $-4\pi\chi$ over 0.15 at 20 K) are not filamentary. Moreover, the results of Rietveld analysis indicate that the $Ca_{1-x}La_xFe_2As_2$ content are 5.1 wt. % at $x_{nom}$ = 0.1, and 0.02 or less wt. % at $x_{nom}$ = 0.2 and 0.3. Such small fractions cannot cause the observed large values (> 0.66) of $-4\pi\chi$. Therefore, we conclude that the observed superconductivity here is not derived from $Ca_{1-x}La_xFeAs_2$ or $Ca_{1-x}La_xFe_2As_2$. Finally we discuss the oxygen incorporated in the samples. This oxygen probably occurs in the hydrogen source during high pressure synthesis, since we dealt with the starting materials inside the glove box filled with Ar gas and used the stainless capsule to isolate the starting materials from an external atmosphere. As already mentioned in the experimental section, $Ca(OH)_2$ in the hydrogen source generates $H_2O$ at ~573K. Although the majority of generated $H_2O$ reacts with $NaBH_4$, a trace of $H_2O$ can be supplied to the samples through the BN separator. The oxygen contamination raises a doubt that superconductivity was maybe caused by the $LaFeAsO_{1-x}H_x$ phase.[32] However, this possibility can be excluded by two experimental results: First, XRD demonstrates that there is no peak derived from $LaFeAsO_{1-x}H_x$. $Ca_{1-x}La_xFeAsH$ and $LaFeAsO_{1-x}H_x$ have the same crystal structure, but we can easily distinguish their peaks, since their peaks do not overlap with each other due to distinct difference in the lattice constants $c$ between CaFeAsH (~825 pm) and LaFeAsO (~875 pm). Second, the observed maximum $T_c$ (~47 K) is significantly higher than that of $LaFeAsO_{1-x}H_x$ (~36 K).

In summary, indirect electron-doped $Ca_{1-x}La_xFeAsH$ was synthesized by the solid state reaction at 1173K under pressure of 2.5GPa. The maximum solubility of La in CaFeAsH was $x \sim 0.3$. Rietveld analysis revealed that La substitutes the Ca sites in the CaH layers and the site occupancy of La equals almost as the nominal $x$. The elemental composition analysis shows that La concentration was comparable to the nominal $x$ and a small amount of oxygen (2-5 mol%) was incorporated in the samples. An anomaly derived from a structural and a magnetic transition in the $\rho$-T curve is suppressed by increasing La-substitution and superconductivity was observed at $x \geq 0.08$. A maximum $T_c$ is 47.4K at $x = 0.23$ where $4\pi\chi$ is close to -1.0. The maximum $T_c$ (47 K) in La-substituted CaFeAsH are much higher than that (23 K) of direct electron doped $CaFe_{1-x}Co_xAsH$. This is the first iron-based 1111-type superconductor except $Ln$FeAsO in which superconductivity is induced by indirect carrier doping.

**Acknowledgment**

The authors would like to thank Silvia Haindl for valuable discussion. This study was supported by the JPSJ FIRST program and MEXT Element Strategy Initiative to form a core organization.
* hosono@lucid.msl.titech.ac.jp

**References**

1) Y. Kamihara, T. Watanabe, M. Hirano, and H. Hosono: J. Am. Chem. Soc. **130**, 3296 (2008).
2) M. Rotter, M. Tegel, and D. Johrendt: Phys. Rev. Lett. **101**, 107006 (2008).
3) C. Gen-Fu, L. Zheng, L. Gang, H. Wan-Zheng, D. Jing, Z. Jun, Z. Xiao-Dong, Z. Ping, W. Nan-Lin, and L. Jian-Lin: Chin. Phys. Lett. **25**, 3403 (2008).
4) G. Wu, H. Chen, T. Wu, Y.L. Xie, Y.J. Yan, R.H. Liu, X.F. Wang, J.J. Ying, and X.H. Chen: J. Phys. Condens. Matter **20**, 422201 (2008).
5) F.-C. Hsu, J.-Y. Luo, K.-W. Yeh, T.-K. Chen, T.-W. Huang, P.M. Wu, Y.-C. Lee, Y.-L. Huang, Y.-Y. Chu, D.-C. Yan, and M.-K. Wu: Proc. Natl. Acad. Sci. **105**, 14262 (2008).
6) J.H. Tapp, Z. Tang, B. Lv, K. Sasmal, B. Lorenz, P.C.W. Chu, and A.M. Guloy: Phys. Rev. B **78**, 060505 (2008).
7) H. Ogino, Y. Matsumura, Y. Katsura, K. Ushiyama, S. Horii, K. Kishio, and J. Shimoyama: Supercond. Sci. Technol. **22**, 075008 (2009).
8) X. Zhu, F. Han, G. Mu, B. Zeng, P. Cheng, B. Shen, and H.-H. Wen: Phys. Rev. B **79**, 024516 (2009).
9) H. Ogino, S. Sato, K. Kishio, J. Shimoyama, T. Tohei, and Y. Ikuhara: Appl. Phys. Lett. **97**, 072506 (2010).
10) J. Guo, S. Jin, G. Wang, S. Wang, K. Zhu, T. Zhou, M. He, and X. Chen: Phys. Rev. B **82**, 180520 (2010).
11) T.P. Ying, X.L. Chen, G. Wang, S.F. Jin, T.T. Zhou, X.F. Lai, H. Zhang, and W.Y. Wang: Sci. Rep. **2**, (2012).
12) H. Takahashi, K. Igawa, K. Arii, Y. Kamihara, M. Hirano, and H. Hosono: Nature **453**, 376 (2008).
13) G.F. Chen, Z. Li, D. Wu, G. Li, W.Z. Hu, J. Dong, P. Zheng, J.L. Luo, and N.L. Wang: Phys. Rev. Lett. **100**, 247002 (2008).
14) Z.-A. Ren, J. Yang, W. Lu, W. Yi, X.-L. Shen, Z.-C. Li, G.-C. Che, X.-L. Dong, L.-L. Sun, F. Zhou, and Z.-X. Zhao: Europhys. Lett. **82**, 57002 (2008).
15) X.H. Chen, T. Wu, G. Wu, R.H. Liu, H. Chen, and D.F. Fang: Nature **453**, 761 (2008).
16) R. Zhi-An, L. Wei, Y. Jie, Y. Wei, S. Xiao-Li, Zheng-Cai, C. Guang-Can, D. Xiao-Li, S. Li-Ling, Z. Fang, and Z. Zhong-Xian: Chin. Phys. Lett. **25**, 2215 (2008).
17) C. de la Cruz, Q. Huang, J.W. Lynn, J. Li, W.R. II, J.L. Zarestky, H.A. Mook, G.F. Chen, J.L. Luo, N.L. Wang, and P. Dai: Nature **453**, 899 (2008).
18) T. Nomura, S.W. Kim, Y. Kamihara, M. Hirano, P.V. Sushko, K. Kato, M. Takata, A.L. Shluger, and H. Hosono: Supercond. Sci. Technol. **21**, 125028 (2008).
19) Y. Qi, Z. Gao, L. Wang, D. Wang, X. Zhang, and Y. Ma: Supercond. Sci. Technol. **21**, 115016 (2008).
20) S. Matsuishi, Y. Inoue, T. Nomura, H. Yanagi, M. Hirano, and H. Hosono: J. Am. Chem.

Soc. **130**, 14428 (2008).

21) M. Tegel, S. Johansson, V. Weiß, I. Schellenberg, W. Hermes, R. Pöttgen, and D. Johrendt: Europhys. Lett. **84**, 67007 (2008).
22) X. Zhu, F. Han, P. Cheng, G. Mu, B. Shen, L. Fang, and H.-H. Wen: Europhys. Lett. **85**, 17011 (2009).
23) Y. Muraba, S. Matsuishi, and H. Hosono: arXiv:1312.5818.
24) P. Cheng, B. Shen, G. Mu, X. Zhu, F. Han, B. Zeng, and H.-H. Wen: Europhys. Lett. **85**, 67003 (2009).
25) G. Wu, Y.L. Xie, H. Chen, M. Zhong, R.H. Liu, B.C. Shi, Q.J. Li, X.F. Wang, T. Wu, Y.J. Yan, J.J. Ying, and X.H. Chen: J. Phys. Condens. Matter **21**, 142203 (2009).
26) S.V. Chong, S. Hashimoto, H. Yamaguchi, and K. Kadowaki: J. Supercond. Nov. Magn. **23**, 1479 (2010).
27) S. Matsuishi, T. Hanna, Y. Muraba, S.W. Kim, J.E. Kim, M. Takata, S. Shamoto, R.I. Smith, and H. Hosono: Phys. Rev. B **85**, 014514 (2012).
28) H. Hosono and S. Matsuishi: Curr. Opin. Solid State Mater. Sci. **17**, 49 (2013).
29) TOPAS, version 4.2 (Bruker AXS, Karlsruhe, Germany, (2009)).
30) T. Nomura, Y. Inoue, S. Matsuishi, M. Hirano, J.E. Kim, K. Kato, M. Takata, and H. Hosono: Supercond. Sci. Technol. **22**, 055016 (2009).
31) T. Hanna, Y. Muraba, S. Matsuishi, N. Igawa, K. Kodama, S. Shamoto, and H. Hosono: Phys. Rev. B **84**, 024521 (2011).
32) S. Iimura, S. Matuishi, H. Sato, T. Hanna, Y. Muraba, S.W. Kim, J.E. Kim, M. Takata, and H. Hosono: Nat. Commun. **3**, 943 (2012).
33) Z.-A. Ren, G.-C. Che, X.-L. Dong, J. Yang, W. Lu, W. Yi, X.-L. Shen, Z.-C. Li, L.-L. Sun, F. Zhou, and Z.-X. Zhao: Europhys. Lett. **83**, 17002 (2008).
34) Z.A. Ren, J. Yang, W. Lu, W. Yi, G.C. Che, X.L. Dong, L.L. Sun, and Z.X. Zhao, (2013).
35) X. Wang, S.R. Ghorbani, G. Peleckis, and S. Dou: Adv. Mater. **21**, 236 (2009).
36) J. Yang, Z.-C. Li, W. Lu, W. Yi, X.-L. Shen, Z.-A. Ren, G.-C. Che, X.-L. Dong, L.-L. Sun, F. Zhou, and Z.-X. Zhao: Supercond. Sci. Technol. **21**, 082001 (2008).
37) C. Wang, L. Li, S. Chi, Z. Zhu, Z. Ren, Y. Li, Y. Wang, X. Lin, Y. Luo, S. Jiang, X. Xu, G. Cao, and Z. Xu: Europhys. Lett. **83**, 67006 (2008).
38) J.-W.G. Bos, G.B.S. Penny, J.A. Rodgers, D.A. Sokolov, A.D. Huxley, and J.P. Attfield: Chem. Commun. 3634 (2008).
39) C.-H. Lee, A. Iyo, H. Eisaki, H. Kito, M.T. Fernandez-Diaz, T. Ito, K. Kihou, H. Matsuhata, M. Braden, and K. Yamada: J. Phys. Soc. Jpn. **77**, 083704 (2008).
40) N. Katayama, K. Kudo, S. Onari, T. Mizukami, K. Sugawara, Y. Sugiyama, Y. Kitahama, K. Iba, K. Fujimura, N. Nishimoto, M. Nohara, and H. Sawa: J. Phys. Soc. Jpn. **82**, 123702 (2013).
41) H. Yakita, H. Ogino, T. Okada, A. Yamamoto, K. Kishio, T. Tohei, Y. Ikuhara, Y. Gotoh, H, Fujihisa, K. Kataoka, H. Eisaki, J. Shimoyama: J. Am. Chem. Soc. **136**, 846 (2014).

42) S.R. Saha, N.P. Butch, T. Drye, J. Magill, S. Ziemak, K. Kirshenbaum, P.Y. Zavalij, J.W. Lynn, and J. Paglione: Phys. Rev. B **85**, 024525 (2012).

Fig. 1. (color online) Crystal structure information of $Ca_{1-x}La_xFeAsH$. (a) Powder XRD patterns of $Ca_{1-x}La_xFeAsH$, with nominal composition ($x_{nom}$) = 0.1 (top left), 0.2 (bottom left), 0.3 (top right) and 0.4 (bottom right). Red (line) and black (+) traces indicate observed and Rietveld-fitted patterns, respectively. The differences between them (blue line) and Bragg positions of the main phase (green bar) are also shown. (b) Lattice parameters $a$ (top) and $c$ (bottom) (c) La contents obtained through Rietveld analysis. (d) As-Fe-As bond angle ($\alpha$).

Fig. 2. (color online) Composition analysis of $Ca_{1-x}La_xFeAsH$. (a) Analyzed $x$ value of $Ca_{1-x}La_xFeAsH$ as function of $x_{nom}$. The value of $x$ was determined by EPMA. (b) Oxygen concentration determined by EPMA (O : smaller blue closed circle and line) and hydrogen concentration determined by TDS (H : green closed triangle and line). The oxygen concentration subtracted by that of $x = 0$ ($y$ : larger blue closed square and line). The total of H and oxygen incorporated in the bulk (H + $y$ : orange closed circle and line).

Fig. 3. (color online) Electrical and magnetic properties of $Ca_{1-x}La_xFeAsH$. (a) $\rho$-$T$ profiles. (b) Zero-field cooling (ZFC) and field cooling (FC) $4\pi\chi$-$T$ curves measured on powdered samples under the magnetic field (H) of 10 Oe.

Fig. 4 (color online) (a) Electronic phase diagram of $Ca_{1-x}La_xFeAsH_{1-y}O_y$ as a function of the total number of doped electron per iron ($N_e = x - y$), superimposed on that of $CaFe_{1-x}Co_xAsH$. (Ref. 23) (b) $T_c$ vs. As-Fe-As bond angle ($\alpha$). The selected $T_c$ values and structural date are those in (Ref. 27, 31-38.)

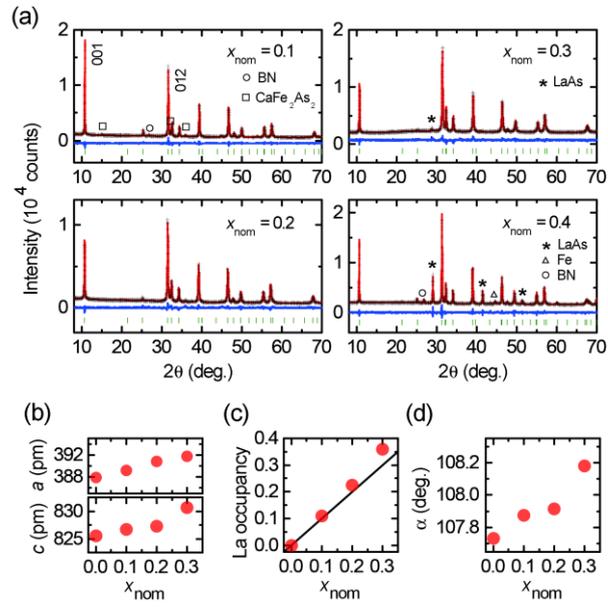

Fig.1.

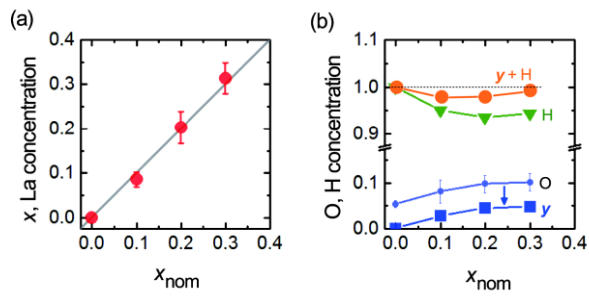

Fig. 2.

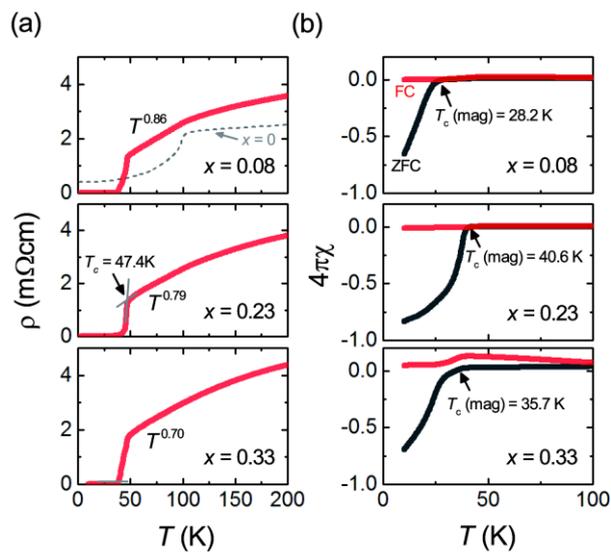

Fig. 3.

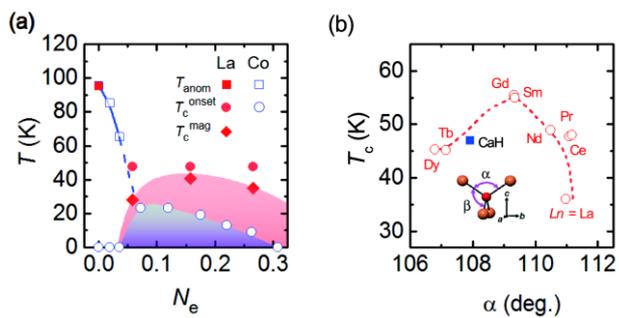

Fig. 4.